\documentclass[a4paper,10pt]{article}

 \usepackage{latexsym}
 \usepackage{amsmath}
 \usepackage{amssymb}
\begin{document}
%\vspace*{6cm}
%%%%%%%%%%%%%%%%%%%%%%%%%%%%%%%%%%%
%
%\begin{flushleft}
%\author{
%P. Kocian\\
%Institut de Physique Th\'eorique, UNI-Lausanne \\
%BSP, 1015 Dorigny, Switzerland
%}
%\end{flushleft}
%

\title{\vspace{-2cm}
\hfill{\normalsize UNIL-IPT/00-13}
\begin{flushleft}
\textbf{Gauge-invariant description\\
of the electromagnetic field\\
in the Higgs phase of the Standard Model}\\
\vspace{1.5cm}
\normalsize{P. Kocian\\
Institut de Physique Th\'eorique, UNI-Lausanne \\
BSP, 1015 Dorigny, Switzerland}
\end{flushleft}
}
\date{}
%\vspace{6cm}
\maketitle
%\vspace{3cm}

\begin{center}
Abstract
\end{center}
It is shown that the definition of the photon field in the broken-symmetry phase of the electroweak
theory proposed recently~\cite{tornkvist2} is equivalent to that proposed previously in Ref.~\cite{shapo}.

\vspace{1.5cm}

\section{Introduction}

In all models of particle physics, electromagnetism appears as part of a larger
symmetry group. For example, in standard electroweak theory the photon is a ``mixture'' of the
$SU(2)$ gauge boson and the hypercharge field; in grand unified theories, the $U(1)$ group is a
subgroup of a simple unifying group, such as $SU(5)$, $SO(10)$, etc. The ordinary
electric ($\vec E$) and magnetic ($\vec B$) fields are directly observable and thus could be
expressed through initial gauge and scalar fields in a gauge-invariant way.
The 't~Hooft expression~\cite{'t hooft} for the electromagnetic tensor for $SO(3)$ gauge theory with
triplet of scalar fields, in which symmetry is broken down to $U(1)$ (Glashow-Georgi model) is a famous example.
This expression is essential for the analysis of the magnetic monopole properties.

This type of analysis has been applied to the case of the electroweak theory~\cite{shapo}, where a
gauge-invariant expression for the electromagnetic field was constructed. Recently, the problem of
gauge-invariant definition of the electromagnetic field in the standard model has attracted a lot of
attention in connection with the possibility of primordial magnetic field generation at the electroweak
phase transition. Such a definition appeared in Ref.~\cite{vachaspati} to show that the
electromagnetic field could be produced, provided the vacuum expectation value of the Higgs field is not constant
during the phase transition. In Ref.~\cite{grasso and riotto}, another definition is used to argue that the magnetic fields may arise 
from some semiclassical configurations of the gauge fields, such as $Z$-strings and
$W$-condensates. However, these definitions~\cite{vachaspati, grasso and riotto} have led to some paradoxes.
For example, they impliy that a magnetic field would always be
present along the internal axis of the electroweak string. This is resolved using the
definition of the electromagnetic tensor~\cite{tornkvist1} that had appeared previously in the study of the electroweak
sphaleron~\cite{hindmarsh} and for the Glashow-Georgi $SO(3)$ model~\cite{'t hooft}. The advantages of that
definition are discussed at length in Refs.~\cite{tornkvist1, tornkvist2}. One of them is that, in the absence of
monopoles, there is no magnetic charge or magnetic current, so there are no
contributions to the electromagnetic tensor from electrically neutral currents.

As well known from the Maxwell theory, an electromagnetic tensor can be defined from a vector
field that undergoes a gauge transformation by the addition of the gradient of an
arbitrary scalar function. Such a vector field was recently constructed from initial scalar and gauge fields 
in the electroweak theory~\cite{tornkvist2}. In this note, we will show that the results of Ref.~\cite{tornkvist2} are in fact
equivalent to the original definition of Ref.~\cite{shapo}.

\section{Construction of the photon field}

With the gauge group $SU(2) \times{} U(1)$, the construction of the photon field given in Ref.~\cite{shapo} may be summarised by choosing a
four-dimensional representation of scalar fields,
\begin{equation}
\boldsymbol{\varphi} = 
\left( \begin{array}{c}
\varphi_{1}\\
\varphi_{2}\\
\varphi_{3}\\
\varphi_{4}\\
\end{array} \right)\ ,
\end{equation}
and the corresponding $SU(2)$ generators :
\begin{equation}
T^{1}=\frac{1}{2}
\left( \begin{array}{cccc}
0 & 0 & 0 & \mathrm{i}\\
0 & 0 & -\mathrm{i} & 0\\
0 & \mathrm{i} & 0 & 0\\
-\mathrm{i} & 0 & 0 & 0
\end{array} \right)\ ,\qquad
T^{2} = \frac{1}{2}
\left( \begin{array}{cccc}
0 & 0 & \mathrm{i} & 0\\
0 & 0 & 0 & \mathrm{i}\\
-\mathrm{i} & 0 & 0 & 0\\
0 & -\mathrm{i} & 0 & 0
\end{array} \right)\ ,\nonumber
\end{equation}
\begin{equation}
T^{3} = \frac{1}{2}
\left( \begin{array}{cccc}
0 & -\mathrm{i} & 0 & 0\\
\mathrm{i} & 0 & 0 & 0\\
0 & 0 & 0 & \mathrm{i}\\
0 & 0 & -\mathrm{i} & 0
\end{array} \right)\ ,
\end{equation}
with respective gauge fields $A^{1}$, $A^{2}$, $A^{3}$ and the $U(1)$ generator
\begin{equation}
Y = \frac{1}{2}
\left( \begin{array}{cccc}
0 & -\mathrm{i} & 0 & 0\\
\mathrm{i} & 0 & 0 & 0\\
0 & 0 & 0 & -\mathrm{i}\\
0 & 0 & \mathrm{i} & 0
\end{array} \right)
\end{equation}
with gauge field $A^{0}$.
The canonical conjugate fields to $\boldsymbol{\varphi}$, $A^{a}_{r}$, $A^{0}_{r}$ are respectively $\mathbf{p}$,
$B^{a}_{r}$, $B^{0}_{r}$. Let us consider new fields given by the canonical transformation
\begin{equation}
(\boldsymbol{\varphi},\ \mathbf{p},\ A^{a}_{r},\ B^{a}_{r},\ A^{0}_{r},\ B^{0}_{r})\ \longrightarrow\ (\rho,\
p_{\rho},\
\theta^{a},\ p_{\theta}^{a},\ \widetilde{A}^{a}_{r},\ \widetilde{B}^{a}_{r},\ \widetilde{A}^{0}_{r},\
\widetilde{B}^{0}_{r})
\end{equation}
where
\begin{equation} \label{4S1}
\boldsymbol{\varphi} = \exp(2\mathrm{i}\  \theta^{a} T^{a})
\left( \begin{array}{c}
0\\
0\\
0\\
\rho
\end{array} \right) \equiv U(\boldsymbol{\varphi})^{\dagger}
\left( \begin{array}{c}
0\\
0\\
0\\
\rho
\end{array} \right), \qquad
\rho = |\boldsymbol{\varphi}|\ ,
\end{equation}
\begin{equation} \label{4S2}
p_{\rho} = \frac{1}{\rho} \ ^{t}\mathbf{p}\ \boldsymbol{\varphi}\ ,
\end{equation}
\begin{equation} \label{4Va1}
\hat{\widetilde{A}}_{r} = U(\boldsymbol{\varphi}) \Big(\widehat{A}_{r}-\frac{1}{\mathrm{i}\ g}\ \partial_{r}\Big)
U(\boldsymbol{\varphi})^{\dagger}\ ,
\end{equation}
\begin{equation} \label{4Va2}
\hat{\widetilde{B}}_{r} = U(\boldsymbol{\varphi}) \ \widehat{B}_{r} \ U(\boldsymbol{\varphi})^{\dagger}\ ,
\end{equation}
\begin{equation} \label {4Vb1}
\widetilde{A}^{0}_{r} = A^{0}_{r}\ ,
\end{equation}
\begin{equation} \label{4Vb2}
\widetilde{B}^{0}_{r} = B^{0}_{r}\ .
\end{equation}
Here, $\widehat{A}_{r}\equiv \sum_{a=1}^{3}A^{a}_{r}T^{a}\ ,\ r=1, 2, 3,$ and
from (\ref{4S1})
\begin{equation} \label{matriceU}
U(\boldsymbol{\varphi})=\frac{1}{\rho}
\left( \begin{array}{cccc}
\varphi_{4} & \varphi_{3} & -\varphi_{2} & -\varphi_{1}\\
-\varphi_{3} & \varphi_{4} & \varphi_{1} & -\varphi_{2}\\
\varphi_{2} & -\varphi_{1} & \varphi_{4} & -\varphi_{3}\\
\varphi_{1} & \varphi_{2} & \varphi_{3} & \varphi_{4}
\end{array} \right)\ .
\end{equation}

The new vector fields are not gauge-invariant :
\begin{eqnarray}
\hat{\widetilde{A}}_{r} & \longrightarrow & \exp(\mathrm{i}\ \lambda_{0}T^{3}) \Big(\hat{\widetilde{A}}_{r} -
\frac{1}{\mathrm{i}\ g}\ \partial_{r} \Big) \exp(-\mathrm{i}\ \lambda_{0} T^{3})\ , \\
\widetilde{A}^{0}_{r} & \longrightarrow & \widetilde{A}^{0}_{r} + \frac{1}{g\prime}\ \partial_{r}\lambda_{0}
\end{eqnarray}
(under a gauge transformation, $\Lambda = \exp(\mathrm{i}\ \lambda^{a} T^{a} + \mathrm{i}\ \lambda_{0} Y)$ with
$\lambda^{1}$, $\lambda^{2}$, $\lambda^{3}$, $\lambda_{0}$ arbitrary real numbers).\\
It can be seen from (12) that the field $\widetilde{A}^{3}$ is only transformed by a
gradient stretching :
\begin{equation}
\widetilde{A}^{3}_{r}\ \longrightarrow\ \widetilde{A}^{3}_{r} + \frac{1}{g}\ \partial_{r}\lambda_{0}\ , \qquad r=1,\ 2,\ 3.
\end{equation}
Thus we have two fields which are transformed by the addition of a pure gradient. The photon field is then given by the
linear combination of $\widetilde{A}^{0}$ and $\widetilde{A}^{3}$,
\begin{equation} \label{photonfield}
E_{r} = \frac{g\ \widetilde{A}^{0}_{r}\ +\ g\prime\ \widetilde{A}^{3}_{r}}{\sqrt{g^{2}+g\prime^{2}}}\ , \qquad r=1,\ 2,\ 3,
\end{equation}
because it transforms by a pure gradient stretching
\begin{equation}
E_{r}\  \longrightarrow\  E_{r} + \frac{1}{e}\ \partial_{r}\lambda_{0}\ , \qquad r=1,\ 2,\ 3,
\end{equation}
where $e=\frac{g\ g\prime}{\sqrt{g^{2}+g\prime^{2}}}$ .

\section{The photon field expressed in terms of the initial fields}

The generators $T^{a},\ a=1,\ 2,\ 3$, satisfy the orthonormalisation condition
$\mathrm{Tr}(T^{a}T^{b})=\delta^{ab}$, so that $\widetilde{A}^{3}_{r}=\mathrm{Tr}(\hat{\widetilde{A}}_{r}T^{3})$
and, from (\ref{4Va1}),
\begin{eqnarray} \label{A3tilde}
\widetilde{A}^{3}_{r} & = & A^{1}_{r}\ \frac{2}{\rho^{2}}\ \big(\varphi_{1}\ \varphi_{3}+\varphi_{2}\
\varphi_{4}\big)+A^{2}_{r}\ \frac{2}{\rho^{2}}\ \big(\varphi_{2}\ \varphi_{3}+\varphi_{1}\ \varphi_{4}\big)+{}
\nonumber\\
& & {}+A^{3}_{r}\ \frac{1}{\rho^{2}}\
\big(-\varphi_{1}^{2}-\varphi_{2}^{2}+\varphi_{3}^{2}+\varphi_{4}^{2}\big)+{}\nonumber\\
& & {}-\frac{1}{g}\ \frac{2}{\rho^{2}}\ \big(\varphi_{1}\ \partial_{r}\varphi_{2}-\varphi_{2}\
\partial_{r}\varphi_{1}+\varphi_{3}\ \partial_{r}\varphi_{4}-\varphi_{4}\ \partial_{r}\varphi_{3}\big)\ .
\end{eqnarray}
Using (\ref{A3tilde}) and (\ref{4Vb1}) in (\ref{photonfield}), $E_{r}$ can be written in terms of the
initial fields :
\begin{eqnarray} \label{photonfield original}
E_{r} & = & \frac{g\ A^{0}_{r}}{\sqrt{g^{2}+g\prime^{2}}}+\frac{g\prime}{\sqrt{g^{2}+g\prime^{2}}}\ \frac{1}{\rho^{2}}\
\Big\{2\big(\varphi_{1}\ \varphi_{3}+\varphi_{2}\ \varphi_{4}\big)\ A^{1}_{r}+{}\nonumber\\
& & {}+2\big(\varphi_{2}\ \varphi_{3}-\varphi_{1}\
\varphi_{4}\big)\ A^{2}_{r}+\big(-\varphi_{1}^{2}-\varphi_{2}^{2}+\varphi_{3}^{2}+\varphi_{4}^{2}\big)\ A^{3}_{r}+{}\nonumber\\
& & {}-\frac{2}{g}\ \big(\varphi_{1}\ \partial_{r}\varphi_{2}-\varphi_{2}\ \partial_{r}\varphi_{1}
+\varphi_{3}\ \partial_{r}\varphi_{4}-\varphi_{4}\ \partial_{r}\varphi_{3}\big)\Big\}\ .
\end{eqnarray}

In Ref.~\cite{tornkvist2}, the following expression is given for the photon field :
\begin{equation} \label{ola tornkvist}
A^{\mathrm{em}}_{\mu}=\cos\theta_{\mathrm{W}}\ A^{0}_{\mu}+\sin\theta_{\mathrm{W}}\Big(-\hat{\phi}^{a}A^{a}_{\mu}+
\frac{\mathrm{i}}{g}\ \mathrm{Tr}(\sigma^{3}\ V^{\dagger}\ \partial_{\mu}V)\Big)
\end{equation}
where $\hat{\phi}^{a}=\frac{\boldsymbol{\Phi}^{\dagger}\sigma^{a}\boldsymbol{\Phi}}{\boldsymbol{\Phi^{\dagger}}\
\boldsymbol{\Phi}}$ , $a$ = 1, 2, 3 ($\sigma^{a}$ being the Pauli spin matrices), 
$V=\frac{1}{\boldsymbol{\Phi}^{\dagger}\ \boldsymbol{\Phi}}
\left( \begin{array}{cc}
\phi_{2}^{\star} & \phi_{1}\\
-\phi_{1}^{\star} & \phi_{2}
\end{array} \right)$
and $\boldsymbol{\Phi}=(\phi_{1},\phi_{2})$ is the complex scalar doublet representation. Defining
\begin{eqnarray} \label{parametrisation}
\phi_{1} & = & \varphi_{1}-\mathrm{i}\ \varphi_{2}\ ,\\
\phi_{2} & = & -\varphi_{3}+\mathrm{i}\ \varphi_{4}\ ,
\end{eqnarray}
we have explicitly
\begin{equation} \label{composante1}
\hat{\phi}^{1}=\frac{1}{\boldsymbol{\Phi}^{\dagger}\ \boldsymbol{\Phi}}\ \big(\phi_{1}^{\star}\ \phi_{2}+\phi_{1}\
\phi_{2}^{\star}\big)=-\frac{2}{\rho^{2}}\ \big(\varphi_{1}\ \varphi_{3}+\varphi_{2}\ \varphi_{4}\big)\ ,
\end{equation}
\begin{equation} \label{composante2}
\hat{\phi}^{2}=\frac{\mathrm{i}}{\boldsymbol{\Phi}^{\dagger}\ \boldsymbol{\Phi}}\ \big(-\phi_{1}^{\star}\ \phi_{2}+\phi_{1}\
\phi_{2}^{\star}\big)=-\frac{2}{\rho^{2}}\ \big(\varphi_{2}\ \varphi_{3}-\varphi_{1}\ \varphi_{4}\big)\ ,
\end{equation}
\begin{equation} \label{composante3}
\hat{\phi}^{3}=\frac{1}{\boldsymbol{\Phi}^{\dagger}\ \boldsymbol{\Phi}}\ \big(\phi_{1}^{\star}\ \phi_{1}-\phi_{2}^{\star}\
\phi_{2}\big)=\frac{1}{\rho^{2}}\ \big(\varphi_{1}^{2}+\varphi_{2}^{2}-\varphi_{3}^{2}-\varphi_{4}^{2}\big)\ ,
\end{equation}
\begin{eqnarray} \label{trace}
\lefteqn{ \mathrm{Tr}\big(\sigma^{3}\ V^{\dagger}\ \partial_{\mu}V\big)=\frac{1}{\boldsymbol{\Phi}^{\dagger}\ \boldsymbol{\Phi}}\ \big(\phi_{2}\
\partial_{\mu}\phi_{2}^{\star}+\phi_{1}\ \partial_{\mu}\phi_{1}^{\star}-\phi_{1}^{\star}\
\partial_{\mu}\phi_{1}-\phi_{2}^{\star}\ \partial_{\mu}\phi_{2}\big){} }\nonumber\\
& & {}=\frac{2\ \mathrm{i}}{\rho^{2}}\ \big(\varphi_{1}\ \partial_{\mu}\varphi_{2}-\varphi_{2}\
\partial_{\mu}\varphi_{1}+\varphi_{3}\ \partial_{\mu}\varphi_{4}-\varphi_{4}\ \partial_{\mu}\varphi_{3}\big)\ .
\end{eqnarray}
Substiting these expressions in (\ref{ola tornkvist}),
\begin{eqnarray}
A^{\mathrm{em}}_{\mu} & = & \frac{g\ A^{0}_{r}}{\sqrt{g^{2}+g\prime^{2}}}+\frac{g\prime}{\sqrt{g^{2}+g\prime^{2}}}\ \frac{1}{\rho^{2}}\
\Big\{2\big(\varphi_{1}\ \varphi_{3}+\varphi_{2}\ \varphi_{4}\big)\ A^{1}_{\mu}+{}\nonumber\\
& & {}+2\big(\varphi_{2}\ \varphi_{3}+\varphi_{1}\ \varphi_{4}\big)\ A^{2}_{\mu}+
\big(-\varphi_{1}^{2}-\varphi_{2}^{2}+\varphi_{3}^{2}+\varphi_{4}^{2}\big)\ A^{3}_{\mu}+{}\nonumber\\
& & {}-\frac{2}{g}\ \big(\varphi_{1}\ \partial_{\mu}\varphi_{2}-\varphi_{2}\ \partial_{\mu}\varphi_{1}+\varphi_{3}\
\partial_{\mu}\varphi_{4}-\varphi_{4}\ \partial_{\mu}\varphi_{3}\big)\Big\}\ .
\end{eqnarray}
This corresponds to our result (\ref{photonfield original}).

\section{Conclusion}

In the Higgs phase of the standard model, it is possible to construct a vector field which,
under a gauge transformation, undergoes the addition
of a pure gradient. The method is that proposed in Ref.~\cite{shapo}, where a four-dimensional real
representation of scalar fields has been used. We then computed the photon field in terms of the initial scalar and gauge fields.
Such an expression has also appeared in Ref.~\cite{tornkvist2},
for a complex scalar doublet representation. We have shown that, using an appropriate parametrisation
of the complex fields by real fields, the two formulations are equivalent.
\\ \\
{\large \textbf{Aknowledgments}}\\
I am grateful to Professor M.E. Shaposhnikov for having proposed this subject to me,
and for his precious help. I thank also Mrs. D. Watson
for helping me to produce this note in good English, and Mr. E. F. Alba for introducing me to \LaTeX.

{\small

\end{document}